# Complex Networks and Epidemiology

## Part I: Complex Networks

James Kim[*]

Center for Applied Mathematics, Institute of Sciences and Technology of Asia

**Abstract**

Complex networks describe a wide range of systems in nature and society. Frequently cited examples include Internet, WWW, a network of chemicals linked by chemical reactions, social relationship networks, citation networks, etc. The research of complex networks has attracted many scientists' attention. Physicists have shown that these networks exhibit some surprising characters, such as high clustering coefficient, small diameter, and the absence of the thresholds of percolation. Scientists in mathematical epidemiology discovered that the threshold of infectious disease disappears on contact networks that following Scale-Free distribution. Researchers in economics and public health also find that the imitation behavior could lead to cluster phenomena of vaccination and un-vaccination. In this note, we will review 1) the basic concepts of complex networks; 2) Basic epidemic models; 3) the development of complex networks and epidemiology.

[*] Chief scientist at Center for Applied Mathematics, Institute of Sciences and Technology of Asia, Hong Kong
Email: j.618.kim@gmail.com

## 1. Introduction

In past few decades, the study of complex network has been one of hottest topics in statistical physics, chemistry, biology, economics, computer science, applied mathematics, and epidemiology [1- 26]. Network, by definition, is a system that contains many individuals that interact with each other through certain rules (see Figure 1). In mathematical terminology, a network is made up of many nodes and links (see Figure 2).

Complex networks can be used to describe the complex relations in nature and society, for example, human contacts, World-Wide Web, movie actor collaboration network, science collaboration graph, the web of human sexual contacts, cellular networks, ecological networks, phone-call network, citation networks, networks in linguistics, power and neural networks, protein folding[1-27]. The research based on complex networks include: the spreading of rumor on complex network, the spreading of infectious disease on social contact networks, computer virus spreading on Internet, the new ideas shared by scientists through citation networks and collaboration networks, human imitation dynamics on social networks[28-43], etc.

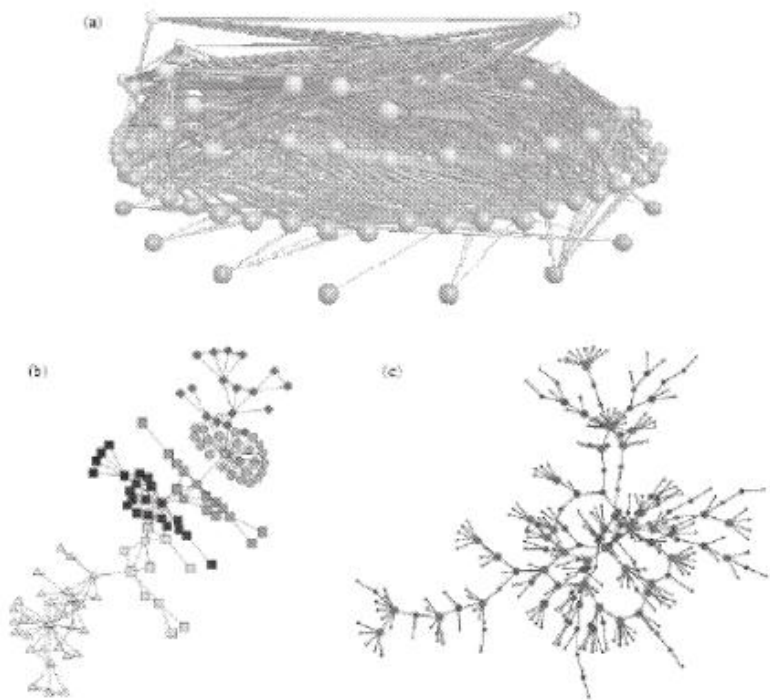

Figure 1: Example of real networks

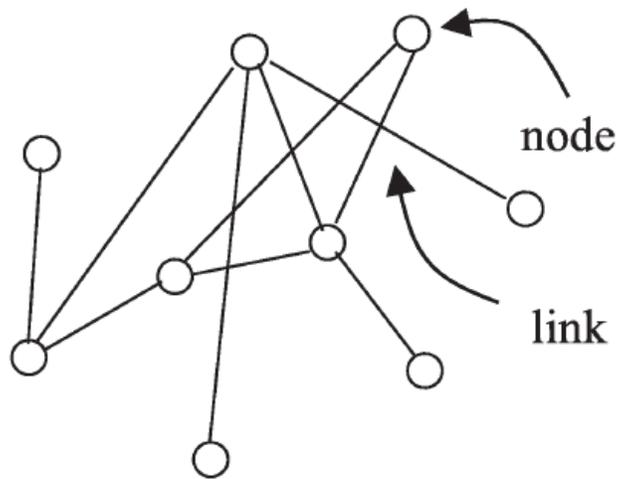

Figure 2: an abstract network

## 2. Basic concepts of complex networks

2.1 Degree

"Degree" is a very basic concept for a node in a complex network. The degree of a given node is defined as the number of links connecting to this node. The degree of a node can also be defined as the number of nearest neighbors of this node [1-10]. Example, in figure 3, degree of node i is 3, while the degree of j is 0.   Not all nodes in a network have the same number of edges. The spread in the number of edges a node has, or node degree, is characterized by a distribution function, which gives the probability that a randomly selected node has exactly k edges. The degree distribution regular lattices have delta-function; degree distribution of random networks follows Poisson distribution. In reality, there are a lot of scale-free networks that have power law degree distribution.  Exponential distribution is another example of degree distribution for networks that have strong degree correlations [27].

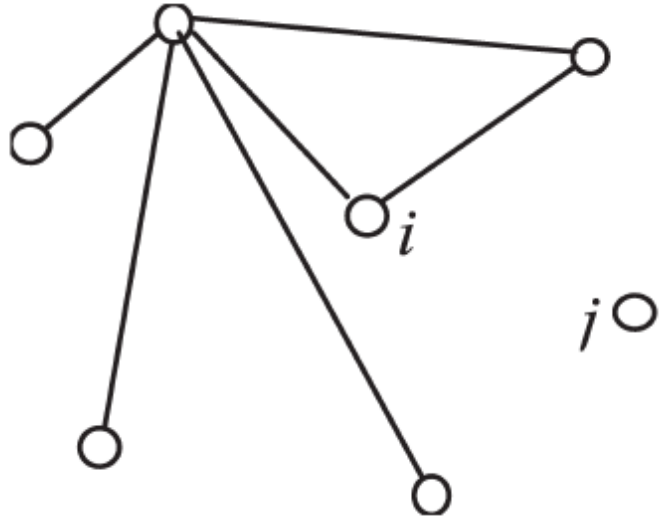

Figure 3: degree of node i is 3. Degree of j is 0.

2.2 Shortest path

In network, there could be many different ways connecting two nodes. In most cases, people are more interested in the "shortest path". The distance between two nodes is defined as the number of links of that the path that connecting these two nodes. If two nodes are next to each other, then the shortest path is 1. The average shortest path for a network is defined as:

$$\bar{l} = \frac{2}{N(N+1)} \sum_{i>j} l_{i,j}$$

(1)

where N is the number of node and $l_{i,j}$ is the distance between nodes i and j.

2.3 Clustering coefficient

For some networks, especially social contact networks, there are cluster phenomena. In order to describe this property, Watts and Strogatz introduced the definition of clustering coefficient:

$$C_i = \frac{2E_i}{k_i(k_i - 1)}$$

(2)

where $k_i$ is the degree of node i and $E_i$ is the number of edges that actually exist between these $k_i$ nodes. The average Clustering coefficient for a network is defined as:

$$C = \frac{1}{n}\sum_i C_i \qquad (3)$$

**3. Complex Network Models**

Scientists have developed various network models based on the complex systems. These models can be used to study the topology of the complex system. On the other hand, these network models provide platform for the research in other fields, for example network design, disease prevention and control.

3.1 The Erdos-Renyi (ER) random network model

ER random network is the simplest network of complex networks [1-7] (see Figure 4). EN network can be constructed as follows:

   a) Generate N nodes in space
   b) Connect any two nodes with a predetermined probability p.

Network properties are summarized below:

   a) Average degree:

$$\langle k \rangle = \frac{2E(N)}{N} = p(N-1) \simeq pN \sim N \qquad (4)$$

b) Degree distribution:

$$p(k) = C_{N-1}^k p^k (1-p)^{N-1-k} \qquad (5)$$

For large N, (5) can be approximated by Poission distribution (see figure 5):

$$p(k) = e^{-pN}\frac{(pN)^k}{k!} = e^{-\langle k \rangle}\frac{\langle k \rangle^k}{k!} \qquad (6)$$

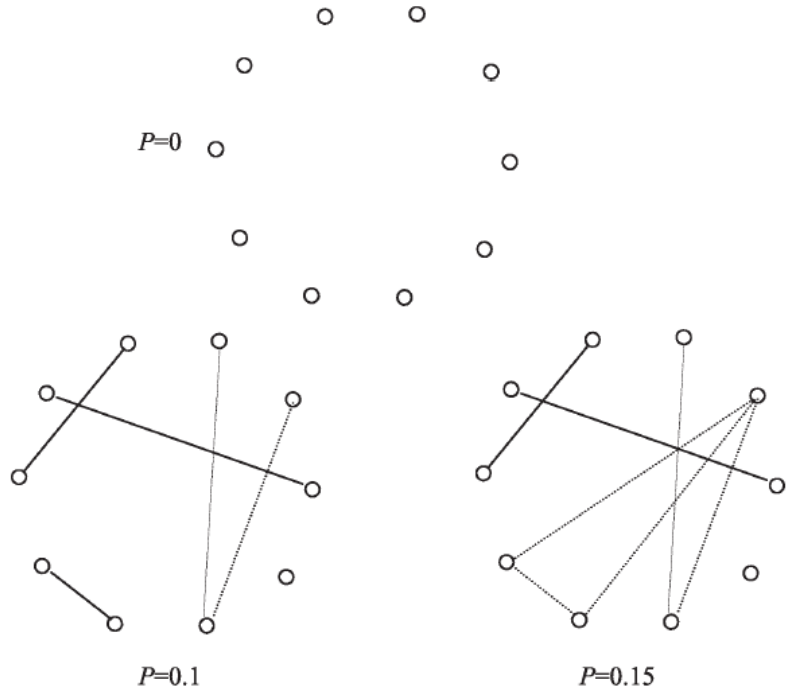

Figure 4: ER network model with N = 10

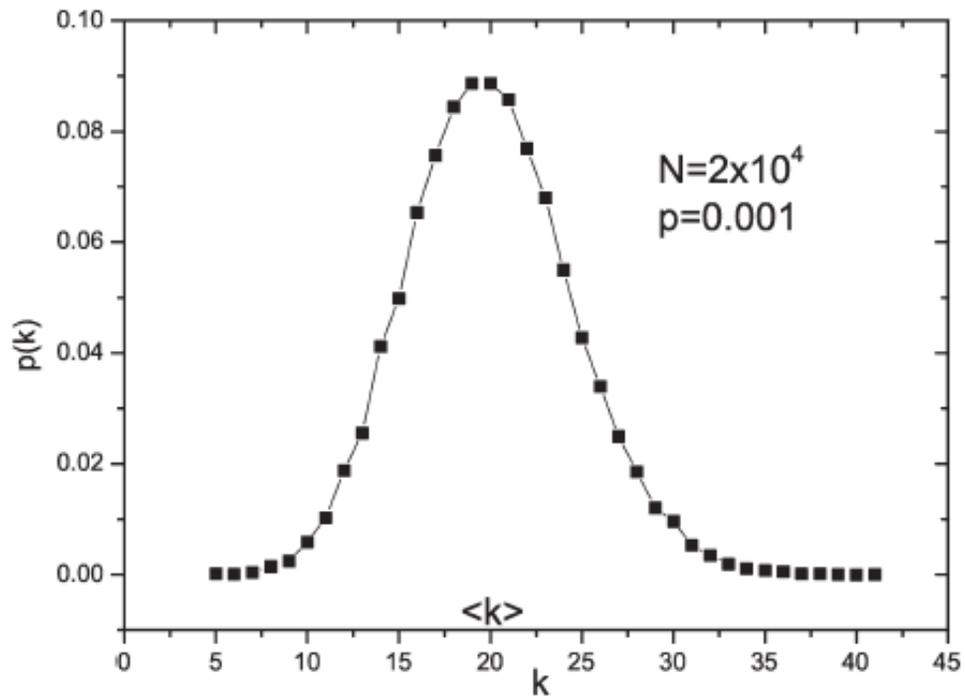

Figure 5: degree distribution of EN network

c) Clustering coefficient:

$$C = p = \frac{\langle k \rangle}{N} \tag{6}$$

d) Average shortest path:

$$\bar{l} \sim \frac{\ln(N)}{\ln(\langle k \rangle)} \tag{7}$$

3.2 Watts-Stogatz Small world network

The small world network has two important properties, large clustering coefficient and smaller average shortest path. The most famous small world network model was developed by Watts and togatz in 1998[16]. There are two steps to construct a Watts-Stogatz Small world network (WS):

a) Start with a ring lattice with N nodes in which every node is connected to its first 2M neighbors (M on either side). In order to have a sparse but connected network at all times, consider N ≫ M ≫ ln(N) ≫ 1.
b) Randomly rewire each edge of the lattice with probability p such that self-connections and duplicate edges are excluded (see Figure6).

Network properties are summarized below:

b) Average degree:

$$<k> = 2M \tag{8}$$

b) Degree distribution is show in figure 7.

c) Clustering coefficient:

$$C = \frac{3(M-1)}{2(2M-1)}(1-p)^3 \tag{9}$$

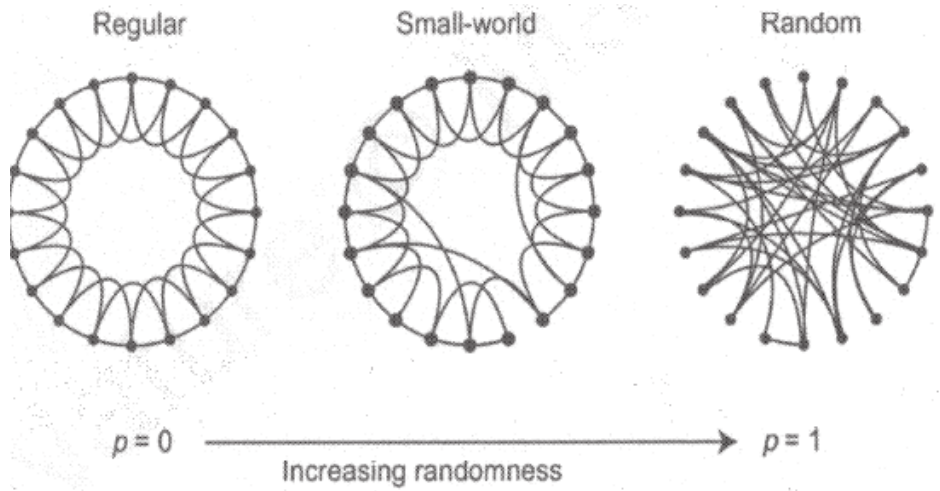

Figure 6: construction of small world networks [16].

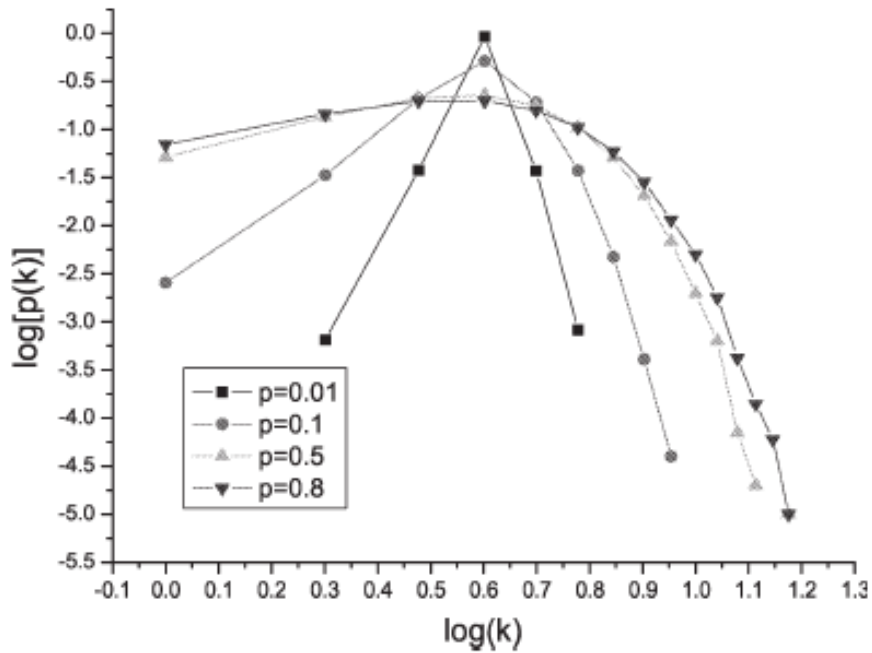

Figure 7: degree distribution of small world network

3.3 The Barabasi-Albert model (Scale-free network)

The Barabasi-Albert model (a.k.a. BA model) introduced in 1998 explains the power-law degree distribution of networks by considering two main ingredients: growth and preferential attachment (Barabasi and Albert 1999). The algorithm used in the BA model goes as follows:

a) Growth: Starting with a small number ($m_0$) of connected nodes, at every time step, we add a new node with m(<$m_0$) edges that link the new node to m different nodes already present in the network.

b) b) Preferential attachment: When choosing the nodes to which the new node connects, we assume that the probability $\pi_i$ that a new node will be connected to node i depends on the degree $k_i$ of node i, such tha

$$\pi_i = \frac{k_i}{\sum_{j \in V} k_j} \tag{10}$$

Scale-free network properties are summarized below:

a) Average degree (see figure 8):

$$<k> = 2m \tag{11}$$

b) Degree distribution:

$$p(k) \sim k^{-\gamma} \tag{12}$$

c) Clustering coefficient:

$$C \sim N^{-0.75} \tag{13}$$

d) Average shortest path:

$$\bar{l} \sim \frac{\ln(N)}{\ln \ln(N)} \tag{14}$$

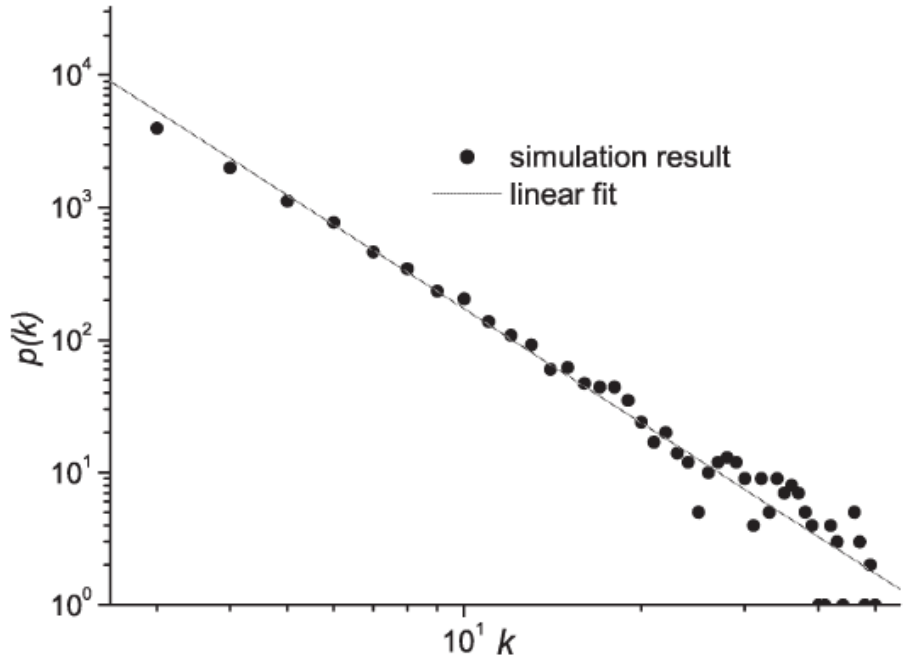

Figure 8: degree distribution of scale-free networks with m = m$_0$ = 3

## 4. Summary

In this review we have discussed the preliminary concepts and methods that are used to describe the networks. We also discuss three fundamental one-level network models, such as ER model and WS small world model. Besides the one-level complex network models, hierarchical network models have also been developed. For example, Liu et al developed a household-structure complex network model [18]. In Liu's model, there is a household at each node. Within a household, family members are fully connected. The disease spreading behavior on the household-structure complex network model is qualitatively different from one-level networks in that there are two infection rates: one is within household and one is among different household [18]. Model details about spread of disease on complex networks will be covered in part II. Following Liu's model, various community structure models have been investigated [44-47]. Because complex networks describe a wide range of systems in nature and society, the research of complex networks will continue to be a hot topic in the future.

**References**

[1] M. E. J. Newman. The structure of scientific collaboration networks. Proc. Natl. Acad. Sci. U.S.A., 98(2):404–409, 2001.

[2] J. H. Fowler and S. Jeon. The authority of Supreme Court precedent. Soc. Netw., 30(1):16–30, 2008.

[3] C. A. Hidalgo, B. Klinger, A.-L. Barab´asi, and R. Hausmann. The product space conditions the development of nations. Science, 317(5837):482–487, 2007.

[4] A. Barrat, M. Barthelemy, R. Pastor-Satorras, and A. Vespignani. The architecture of complex weighted networks. Proc. Natl. Acad. Sci. U.S.A., 101(11):3747–3752, 2004.

[5] M. Kivel¨a, R. K. Pan, K. Kaski, J. Kert´esz, J. Saram¨aki, and M. Karsai. Multiscale analysis of spreading in a large communication network. J. Stat. Mech., page P03005, 2012.

[6] X. Sun, Y. Liu, B. Li, J. Han, and X. Liu. Mathematical model for spreading dynamics of social network worms. J. Stat. Mech., page P04009, 2012.

[7] G. Kossinets and D. J. Watts. Empirical analysis of an evolving social network. Science, 311(5757):88–90, 2006.Formation of structure in growing networks 13

[8] J. P. Onnela, A. Chakraborti, K. Kaski, J. Kert´esz, and A. Kanto. Dynamics of market correlations: Taxonomy and portfolio analysis. Phys. Rev. E, 68(5):056110, 2003.

[9] Sergi Valverde, Ricard V. Sol´e, Mark A. Bedau, and Norman Packard. Topology and evolution of technology innovation networks. Phys. Rev. E, 76(5):056118, 2007.

[10] R. Albert and A.-L. Barab´asi. Statistical mechanics of complex networks. Rev. Mod. Phys., 74(1):47–97, 2002.

[11] M. E. J. Newman and M. Girvan. Finding and evaluating community structure in networks. Phys. Rev. E, 69(2):026113, 2004.

[12] M. Mitzenmacher. A brief history of generative models for power law and lognormal distributions. Internet Mathematics, 1(2):226–251, 2004.

[13] A.-L. Barab´asi and R. Albert. Emergence of scaling in random networks. Science, 286(5439):509–512, 1999.

[14] H. Louch. Personal network integration: Transitivity and homophily in strong-tie relations. Soc.Netw., 22(1):45–64, 2000.

[15] M. E. J. Newman. The structure and function of complex networks. SIAM Rev., 45(2):167–256, 2003.

[16] D. J. Watts and S. H. Strogatz. Collective dynamics of 'small-world' networks. Nature, 393(6684):440–442, 1998.

[17] Liu, J., Tang, Y., and Yang, Z.R., 2004a. The spread of disease with birth and death on networks. J. Stat. Mech., P08008.

[18] Liu, J., Wu, J., Yang, Z.R., 2004b. The spread of infectious disease on complex networks with household-structure. Physica A 341, 273–280.

[19] Xiang Li and Guanrong Chen. A local-world evolving network model. Physica A, 328(1-2):274–286, 2003.

[20] P. Holme and B. J. Kim. Growing scale-free networks with tunable clustering. Phys. Rev. E, 65(2):026107, 2002.

[21] Z. Zhang, L. Rong, B. Wang, S. Zhou, and J. Guan. Local-world evolving networks with tunable clustering. Physica A, 380:639–650, 2007.

22] M. O. Jackson and B. W. Rogers. Meeting strangers and friends of friends: How random are social networks? Am. Econ. Rev., 97(3):890–915, 2007.

[23] Z. Shao, X. Zou, Z. Tan, and Z. Jin. Growing networks with mixed attachment mechanisms. J. Phys. A, 39(9):2035–2042, 2006.

[24] P. Moriano and J. Finke. Power-law weighted networks from local attachments. Europhys. Lett., 99(1):18002, 2012.

[25] G´abor Cs´anyi and Bal´azs Szendr˝oi. Structure of a large social network. Phys. Rev. E, 69(3):036131, 2004.

[26] A. Papoulis. Probability, Random Variables, and Stochastic Processes. McGraw Hill Higher Education, 4th edition, 2002.

[27] Liu J Z and Tang Y F 2005 Chin. Phys. 14 643.

[28] Mbah M, Liu J, Bauch C, Tekel Y, Medlock J, Meyers L and Galvani A 2012 PLoS Computational Biology 8 e1002469.

[29] Epstein J, Parker J, Cummings D and Hammond R 2008 PLoS One 3 e3955

[30] Wang Y, Xiao G, Wong L, Fu X, Ma S and Cheng T 2011 Journal of Physics A: Mathematical and Theoretical 44 355101

[31] d'Onofrio A, Manfredi P and Salinelli E 2007 Theoretical population biology 71 301–317

[32] Zhang H, Small M, Fu X, Sun G and Wang B 2012 Physica D: Nonlinear Phenomena

[33] Chen F 2009 Mathematical biosciences 217 125–133

[34] Chen F, Jiang M, Rabidoux S and Robinson S 2011 Journal of Theoretical Biology

[35] Zhang H, Zhang J, Zhou C, Small M and Wang B 2010 New Journal of Physics 12 023015

[36] Perc M 2012 Scientific Reports 2

[37] Szolnoki A and Perc M 2011 Physical Review E 84 047102

[38] Salath´e M and Bonhoeffer S 2008 Journal of The Royal Society Interface 5 1505–1508

[39] Eames K 2009 Journal of The Royal Society Interface 6 811–814

[40] Barab´asi A and Albert R 1999 science 286 509–512

[41] Erd˝os P and R´enyi A 1960 On the evolution of random graphs (Akad. Kiad´o)

[42] Dybiec B, Kleczkowski A and Gilligan C 2004 Physical Review E 70 066145

[43] Perra N, Balcan D, Gon¸calves B and Vespignani A 2011 PloS one 6 e23084

[44] Sun H.J., Gao Z.Y., Dynamical behaviors of epidemics on scale-free networks with community structure, Physica A: Statistical Mechanics and its Applications, Volume 381, 15 July 2007, Pages 491-496, ISSN 0378-4371

[45] Pautasso M., Xu X., Jeger M. J., Harwood T. D., Moslonka-Lefebvre M. and Pellis L. (2010), Disease spread in small-size directed trade networks: the role of hierarchical categories. Journal of Applied Ecology, 47: 1300–1309. doi: 10.1111/j.1365-2664.2010.01884.x.

[46] Zhang J, Zhen J, Epidemic spreading on complex networks with community structure, Applied Mathematics and Computation, Volume 219, Issue 6, 25 November 2012, Pages 2829-2838.

[47] Ni S, Weng W, Shen S, Fan W, Epidemic outbreaks in growing scale-free networks with local structure, Physica A: Statistical Mechanics and its Applications, Volume 387, Issue 21